\documentclass[aps,prl,reprint,superscriptaddress,preprintnumbers,amsmath,amssymb]{revtex4-2}

\usepackage{amsfonts}
\usepackage{amsmath}
\usepackage{amssymb}
\usepackage{amstext}
\usepackage{bm}
\usepackage{booktabs}
\usepackage{braket}
\usepackage{color}%
\usepackage{dcolumn}
\usepackage{dsfont}
\usepackage{graphicx}
\usepackage{mathrsfs}
\usepackage{mathtools}
\usepackage{mleftright}
\usepackage{needspace}
\usepackage{times}
\usepackage{txfonts}
\usepackage{iftex}
\ifxetex
\DeclareFontFamily{TU}{txr}{}
\DeclareFontShape{TU}{txr}{m}{it}{<->ssub*lmr/m/it}{}
\DeclareFontShape{TU}{txr}{b}{n}{<->ssub*lmr/b/n}{}
\DeclareFontShape{TU}{txr}{bx}{n}{<->ssub*lmr/bx/n}{}
\fi
\usepackage[utf8]{inputenc}

\usepackage{hyperref}
\hypersetup{colorlinks=true,linkcolor=blue,citecolor=blue,urlcolor=blue}

\setcitestyle{open={[},close={]},citesep={,\!},numbers}

\newcommand{\RMP}{Rev.\ Mod.\ Phys.}
\newcommand{\PRL}{Phys.\ Rev.\ Lett.}
\newcommand{\PRA}{Phys.\ Rev.\ A}
\newcommand{\PRB}{Phys.\ Rev.\ B}
\newcommand{\PRE}{Phys.\ Rev.\ E}
\newcommand{\PRResearch}{Phys.\ Rev.\ Research}
\newcommand{\PRAppl}{Phys.\ Rev.\ Applied}

\newcommand{\NJP}{New J.\ Phys.}
\newcommand{\AQT}{Adv.\ Quantum Technol.}
\newcommand{\NatPhys}{Nat.\ Phys.}
\newcommand{\EPL}{Europhys.\ Lett.}
\newcommand{\CMP}{Commun.\ Math.\ Phys.}
\newcommand{\JSP}{J.\ Stat.\ Phys.}

\newcommand{\PhysScr}{Phys.\ Scr.}

\begin{document}
\setlength{\abovedisplayskip}{6pt plus 1pt minus 3pt}
\setlength{\belowdisplayskip}{6pt plus 1pt minus 3pt}
\setlength{\abovedisplayshortskip}{4pt plus 1pt minus 3pt}
\setlength{\belowdisplayshortskip}{4pt plus 1pt minus 3pt}
\setlength{\textfloatsep}{5pt plus 1pt minus 2pt}
\setlength{\dbltextfloatsep}{5pt plus 1pt minus 2pt}
\setlength{\floatsep}{5pt plus 1pt minus 2pt}
\setlength{\abovecaptionskip}{6pt}

\title{Extractable Work from Collective Dark-State Charging of Quantum Batteries}

\author{Jian-Jian Cheng}
\email{chengjianjian@xupt.edu.cn}
\affiliation{School of Science, Xi'an University of Posts and Telecommunications, Xi'an 710121, China}

\author{Ming-Liang Hu}
\email{mingliang0301@163.com}
\affiliation{School of Science, Xi'an University of Posts and Telecommunications, Xi'an 710121, China}

\begin{abstract}
We consider two ensembles of superconducting qubits coupled to a common lossy resonator, one serving as the charger and the other the battery. Destructive interference between their collective decay amplitudes selects a tunable family of many-body dark states, allowing controlled excitation transfer from the charger to the battery. The collective charging dynamics preserves the symmetric Dicke subspace of the battery, so its reduced state has at most $N+1$ nonzero eigenvalues. Together with the $N$-fold degeneracy of the battery's first excited energy level, this restriction keeps the passive energy of the entire battery below the single-qubit excitation energy $\omega_0$, independently of the battery size. When the average stored energy per battery qubit is fixed and nonzero, the total stored energy grows linearly with $N$, while the ratio of global ergotropy to stored energy approaches unity as $N$ increases.
\end{abstract}

\maketitle

\looseness=1\textit{Introduction.---}Collective charging offers a way to improve the performance of quantum batteries \cite{ref1,ref3,ref72,ref76,ref78}. Previous studies have shown that collective interactions can provide a charging-power advantage over independent charging \cite{ref4,ref5,ref74,ref75,ref91,ref15} and enable efficient energy transfer between a charger and a battery \cite{ref6,ref7,ref73,ref17,ref18}. However, not all the energy stored in a battery can be extracted as work \cite{ref2}. For a given battery Hamiltonian, the extractable work depends not only on the stored energy but also on the state of the battery \cite{ref13}. Thus, besides charging power and stored energy, the amount of energy available as work is an important measure of battery performance \cite{ref79,ref9,ref81}.

The role of quantum states in work extraction has been widely studied. Previous work has examined how correlations between the charger and battery, as well as correlations among battery qubits, affect charging and work extraction \cite{ref8,ref77}. The effects of quantum coherence and entanglement on extractable work have also been investigated \cite{ref12,ref49,ref88}. These studies show that batteries with the same stored energy can yield different amounts of work. Increasing the stored energy alone therefore does not guarantee a corresponding increase in extractable work. The state prepared during charging also plays an important role in determining how much work can be extracted.

\looseness=1Previous studies of dissipation in quantum batteries have addressed state preparation and control \cite{ref92,ref85,ref84,ref25,ref26,ref27}, work extraction \cite{ref20,ref86,ref21,ref83}, dark-state charging and stabilization \cite{ref24,ref87,ref93}, and collective decay through a common environment \cite{ref29,ref90,ref28,ref89}. For independently charged qubits, the passive energy grows linearly with $N$ when each qubit is prepared in the same mixed state as $N$ increases. The passive energy depends on the eigenvalues of the battery state and the energy spectrum of its Hamiltonian, rather than on the stored energy alone.

\looseness=1Here we consider two ensembles of superconducting qubits coupled to a common lossy resonator, one serving as the charger and the other as the battery. Interference between their collective decay amplitudes selects a family of many-body dark states connecting the initially excited charger state to the fully charged battery state. Varying the relative coupling strengths allows excitation transfer along this dark-state family. The collective dynamics preserves the symmetric Dicke subspace of the battery, so its reduced state has at most $N+1$ nonzero eigenvalues. In the corresponding passive state, these eigenvalues occupy the ground state and the $N$-fold degenerate first excited level of the battery Hamiltonian. Consequently, the passive energy of the entire battery remains below the single-qubit excitation energy $\omega_0$, independently of the battery size. When the average stored energy per battery qubit remains fixed and nonzero, the total stored energy grows linearly with $N$, while the ratio of global ergotropy to stored energy approaches unity as $N$ increases. Numerical simulations at finite sweep rates also show that substantial stored energy and global ergotropy are retained over the range of parameters studied.

\begin{figure}[t]
\centering
\includegraphics[width=\columnwidth]{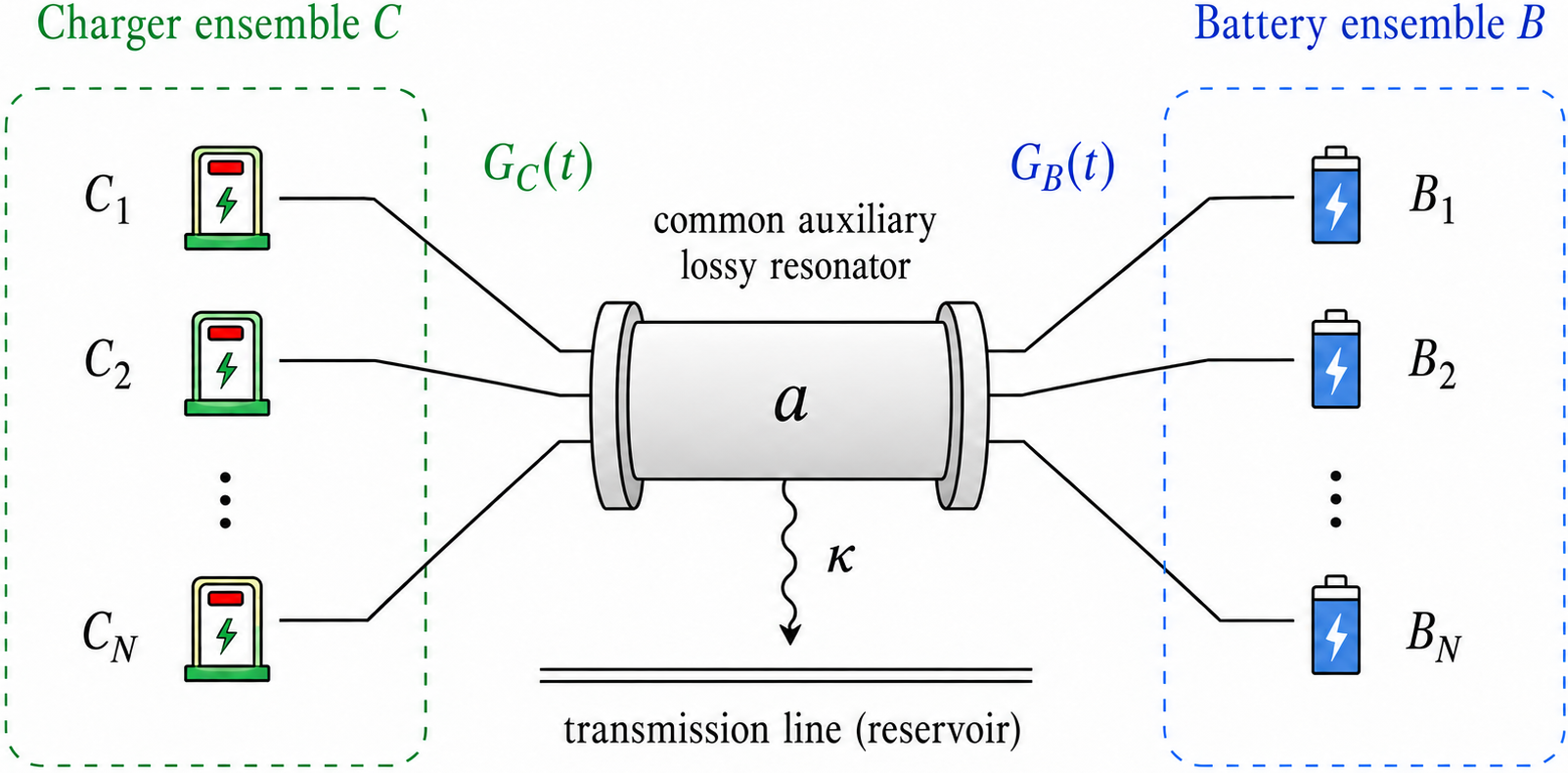}
\caption{Circuit-QED architecture for collective dissipative charging. Two ensembles of $N$ transmon qubits, serving as charger $C$ and battery $B$, couple with effective time-dependent amplitudes $G_C(t)$ and $G_B(t)$ to a common auxiliary lossy resonator $a$. The resonator is coupled to a transmission line, which acts as a photon-loss reservoir at rate $\kappa$. In the bad-cavity regime, eliminating $a$ yields the collective dissipative dynamics of Eq.~\eqref{eq:2}.} \label{fig:circuit_qed_architecture}
\end{figure}

\textit{Circuit-QED collective channel.---}
As shown in Fig.~\ref{fig:circuit_qed_architecture}, we consider two ensembles of superconducting qubits, a charger $C=\{C_1,\ldots,C_N\}$ and a battery $B=\{B_1,\ldots,B_N\}$, each consisting of $N$ transmons. We restrict each transmon to its two lowest energy levels. All qubits have transition frequency $\omega_0$ and couple to a common auxiliary resonator of frequency $\omega_r$, with annihilation operator $a$ and photon-loss rate $\kappa$. We set $\hbar=1$ and work in the rotating frame of the bare qubits and resonator.
Parametric control of the qubit--resonator couplings \cite{ref36,ref35,ref37} activates resonant and phase-controlled exchange between each qubit and the common resonator \cite{ref34,ref39},
\begin{equation}
H_{\rm int}(t)
=
\sum_{\alpha=C,B}\sum_{j=1}^{N}
\left[
G_\alpha(t)e^{i\varphi_\alpha}
a^\dagger\sigma^-_{\alpha j}
+\mathrm{H.c.}
\right].
\label{eq:Hint}
\end{equation}
Here $G_\alpha(t)$ sets the exchange amplitude and $\varphi_\alpha$ its phase. The operator $\sigma_{\alpha j}^-=|g\rangle_{\alpha j}\langle e|$ lowers the $j$th qubit of ensemble $\alpha$. Equal controls within each ensemble give the collective lowering operators $S_\alpha^-=\sum_{j=1}^{N}\sigma_{\alpha j}^-\quad(\alpha=C,B)$. Since both ensembles couple coherently to the same resonator mode, the amplitudes for creating a resonator excitation interfere before that excitation is lost to the transmission line. The interference depends only on the relative phase between the two ensemble--resonator couplings. We choose the phase difference $\varphi_B-\varphi_C=\pi$, so that the two collective amplitudes enter with opposite signs. With this phase choice, the unreduced qubit--resonator dynamics is
\begingroup
\begin{equation} \label{eq:1}
\begin{aligned}
\dot\rho_{\rm full}
&=
-i\!\left[
H_{\rm full}(t),\rho_{\rm full}
\right]
+\kappa\mathcal D[a]\rho_{\rm full},
\\
H_{\rm full}(t)
&=
a^\dagger F_N(t)+F_N^\dagger(t)a,
\\
F_N(t)
&=
G_C(t)S_C^- -G_B(t)S_B^- .
\end{aligned}
\end{equation}
\endgroup
Here $\mathcal D[O]\rho=O\rho O^\dagger-\{O^\dagger O,\rho\}/2$ is the Lindblad dissipator, and $F_N(t)$ is a collective lowering operator acting on the two qubit ensembles. The terms $a^\dagger S_C^-$ and $S_B^+a$ describe excitation exchange between the resonator and the charger or battery, respectively. The resonator thus mediates energy transfer between the two ensembles.

In the bad-cavity regime, we require $\kappa\gg\max_t\|F_N(t)\|$, where $\|\cdot\|$ denotes the operator norm. For each ensemble of $N$ qubits, $\|S_\alpha^-\|\leq(N+1)/2$, giving $\|F_N(t)\|\leq(N+1)[|G_C(t)|+|G_B(t)|]/2$. We also assume that the couplings vary slowly on the cavity relaxation timescale $1/\kappa$. Adiabatically eliminating the resonator \cite{ref45}, we obtain
\begin{equation} \label{eq:2}
\begin{aligned}
\dot\rho
&=
\mathcal D[L_N(t)]\rho,
\\
L_N(t)
&=
\sqrt{\Gamma_C(t)}\,S_C^-
-
\sqrt{\Gamma_B(t)}\,S_B^-,
\\
\Gamma_\alpha(t)
&=
\frac{4G_\alpha^2(t)}{\kappa}
\qquad
(\alpha=C,B).
\end{aligned}
\end{equation}
Here $\rho(t)$ describes the two qubit ensembles after eliminating the resonator, whereas $\rho_{\rm full}(t)$ in Eq.~\eqref{eq:1} includes the resonator. Since both ensembles couple to the same lossy mode, the effective dynamics contains a single collective jump operator $L_N(t)$, rather than two independent decay channels. The two decay amplitudes can therefore interfere destructively, giving rise to collective dark states.

We use the bare ensemble Hamiltonians $H_\alpha=\omega_0\hat n_\alpha$, where $\hat n_\alpha=\sum_{j=1}^{N}\sigma_{\alpha j}^{+}\sigma_{\alpha j}^{-}$ counts the excitations in ensemble $\alpha=C,B$. The energy stored in the battery is $E_B(t)=\mathrm{Tr}[\rho(t)H_B]$. We quantify the work extractable from the battery by its global ergotropy \cite{ref46,ref47,ref48,ref50,ref80},
\begin{equation} \label{eq:ergotropy}
\mathcal W_B(t)
=
E_B(t)
-
\min_{U_B}
\operatorname{Tr}
\!\left[
U_B\rho_B(t)U_B^\dagger H_B
\right].
\end{equation}
Here $\rho_B(t)=\operatorname{Tr}_C\rho(t)$, and the minimization is over all unitaries $U_B$ on the full $2^N$-dimensional battery Hilbert space. The minimum is the passive energy $E_{\rm pas}(t)$. A state with zero ergotropy is passive.

\begin{figure*}[!t]
\centering
\includegraphics[width=\textwidth]{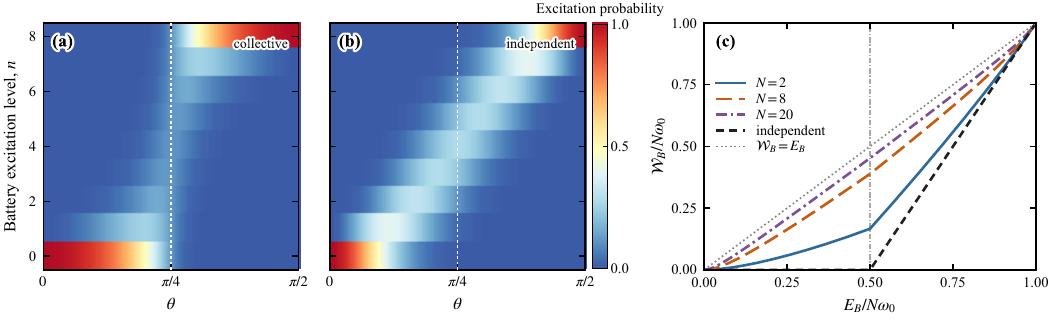}
\caption{Battery excitation distributions and global ergotropy. (a) Probability $p_n^{\mathrm D}$ of finding $n$ excitations in the battery for the collective dark state with $N=8$, as a function of $\theta$, defined by $\tan\theta=G_C/G_B$. (b) Corresponding probability $p_n^{\mathrm{ind}}$ for $N=8$ independent charger--battery pairs. The vertical dashed lines mark $\theta=\pi/4$, where $p_n^{\mathrm D}=1/(N+1)$ and $p_n^{\mathrm{ind}}=\binom{N}{n}/2^N$. (c) Global ergotropy $\mathcal{W}_B/N\omega_0$ versus stored energy $E_B/N\omega_0$ for $N=2$, $8$, and $20$. The dashed curve shows the independent-pair result, the dotted line marks $\mathcal{W}_B=E_B$, and the vertical line indicates half filling.} \label{fig:collective_battery}
\end{figure*}

\textit{Collective dark-state charging.---}
We initialize the system in $|N,0\rangle=|e\rangle_C^{\otimes N}|g\rangle_B^{\otimes N}$, with all charger qubits \mbox{excited} and all battery qubits in their ground states. The target state $|0,N\rangle=|g\rangle_C^{\otimes N}|e\rangle_B^{\otimes N}$ has the opposite occupations: a fully discharged charger and a fully charged battery. Because the qubits within each ensemble couple identically to the resonator, the dynamics is invariant under permutations of qubits within that ensemble. Since the initial state is symmetric under these permutations, the evolution remains within the symmetric Dicke subspace of each ensemble. We seek dark states in the sector with $N$ total excitations, which contains both the initial state $|N,0\rangle$ and the target state $|0,N\rangle$. A convenient basis is $|N-n,n\rangle\equiv|N-n\rangle_C\otimes|n\rangle_B$ $(n=0,\ldots,N)$, where $|k\rangle_\alpha$ denotes the normalized symmetric Dicke state with $k$ excitations in ensemble $\alpha$. A general state can then be written as $|\psi_N\rangle=\sum_{n=0}^{N}d_n|N-n,n\rangle$, where $d_n$ is the probability amplitude of the corresponding Dicke basis state. A dark state does not emit through the collective decay channel. We seek dark states satisfying $L_N|\psi_N\rangle=0$, or equivalently $F_N|\psi_N\rangle=0$. For $n=0,\ldots,N-1$, the state $|N-n-1,n\rangle$ can be reached by lowering the charger in $|N-n,n\rangle$ or the battery in $|N-n-1,n+1\rangle$:
\begin{equation} \label{eq:paths}
\begin{aligned}
&S_C^-|N-n,n\rangle
=
\sqrt{(N-n)(n+1)}
\,|N-n-1,n\rangle ,
\\
&S_B^-|N-n-1,n+1\rangle
=
\sqrt{(n+1)(N-n)}
\,|N-n-1,n\rangle .
\end{aligned}
\end{equation}
Both processes leave the two ensembles in the same final Dicke state and couple to the same resonator mode, so their emission amplitudes are indistinguishable and can interfere. Using Eq.~\eqref{eq:paths} to evaluate $F_N|\psi_N\rangle$, we obtain

\begingroup
\footnotesize
\setlength{\arraycolsep}{1.5pt}
\begin{equation} \label{eq:matrix}
\begin{gathered}
F_N|\psi_N\rangle
=
\begin{pmatrix}
|N-1,0\rangle&
|N-2,1\rangle&
\cdots&
|0,N-1\rangle
\end{pmatrix}
\\[0.5mm]
\times\;
\begin{pmatrix}
G_C\sqrt{N} & -G_B\sqrt{N} & 0 & \cdots & 0\\
0 & G_C\sqrt{2(N\!-\!1)} & -G_B\sqrt{2(N\!-\!1)} & \ddots & \vdots\\
\vdots & \ddots & \ddots & \ddots & 0\\
0 & \cdots & G_C\sqrt{2(N\!-\!1)} & -G_B\sqrt{2(N\!-\!1)} & 0\\
0 & \cdots & 0 & G_C\sqrt{N} & -G_B\sqrt{N}
\end{pmatrix}
\\[0.5mm]
\times\;
\begin{pmatrix}
d_0& d_1& \cdots& d_N
\end{pmatrix}^{\mathrm{T}}
\end{gathered}
\end{equation}
\endgroup
The $N\times(N+1)$ matrix in Eq.~\eqref{eq:matrix} represents $F_N$ between the symmetric subspaces with $N$ and $N-1$ total excitations. Since the final Dicke states are orthogonal, $F_N|\psi_N\rangle=0$ requires each coefficient to vanish, giving
\begingroup
\setlength{\abovedisplayskip}{5pt plus 1pt minus 3pt}%
\setlength{\belowdisplayskip}{5pt plus 1pt minus 3pt}%
\setlength{\abovedisplayshortskip}{3pt plus 1pt minus 3pt}%
\setlength{\belowdisplayshortskip}{3pt plus 1pt minus 3pt}%
\begin{equation} \label{eq:recurrence}
G_Cd_n=G_Bd_{n+1}
\qquad
(n=0,\ldots,N-1).
\end{equation}
\endgroup
For $G_CG_B\neq0$, Eq.~\eqref{eq:recurrence} gives $d_{n+1}/d_n=G_C/G_B$, independent of $n$.  Its kernel is therefore one dimensional, so the dark state is unique up to normalization and an overall phase. Introducing $\tan\theta=G_C/G_B=\sqrt{\Gamma_C/\Gamma_B}$ and normalizing the solution of Eq.~\eqref{eq:recurrence}, we obtain the dark state
\begingroup
\setlength{\abovedisplayskip}{5pt plus 1pt minus 3pt}%
\setlength{\belowdisplayskip}{5pt plus 1pt minus 3pt}%
\setlength{\abovedisplayshortskip}{3pt plus 1pt minus 3pt}%
\setlength{\belowdisplayshortskip}{3pt plus 1pt minus 3pt}%
\begin{equation} \label{eq:dark_state}
\begin{aligned}
|D_N(\theta)\rangle
&=
\frac{1}{\sqrt{Z_N(\theta)}}
\sum_{n=0}^{N}
\cos^{N-n}\theta\,
\sin^{n}\theta
|N-n,n\rangle ,
\\
Z_N(\theta)
&=
\sum_{n=0}^{N}
\cos^{2(N-n)}\theta\,
\sin^{2n}\theta .
\end{aligned}
\end{equation}
\endgroup
The dark state is $|N,0\rangle$ at $\theta=0$ and $|0,N\rangle$ at $\theta=\pi/2$. As $\theta$ increases between these limits, the mean number of battery excitations increases from zero to $N$. When the couplings vary in time, however, the instantaneous dark state is not an exact solution of Eq.~\eqref{eq:2}. Remaining close to this state requires the sweep to be slow compared with the relevant dissipative relaxation rates \cite{ref51,ref52}.

Since the charger Dicke states are orthogonal, tracing out the charger removes the terms with different excitation numbers. The reduced battery state is therefore diagonal in the symmetric Dicke basis,
\begin{equation} \label{eq:reduced_battery}
\begin{gathered}
\rho_B^{\mathrm D}
=
\operatorname{Tr}_C|D_N\rangle\langle D_N|
=
\sum_{n=0}^{N}
p_n^{\mathrm D}
|n\rangle_B\langle n|,
\\
p_n^{\mathrm D}
=
\frac{\cos^{2(N-n)}\theta\,\sin^{2n}\theta}{Z_N(\theta)} ,
\end{gathered}
\end{equation}
The states $|n\rangle_B$ are orthonormal, so $p_n^{\mathrm D}$ are the nonzero eigenvalues of $\rho_B^{\mathrm D}$ for $0<\theta<\pi/2$. From Eq.~\eqref{eq:reduced_battery}, $p_{n+1}^{\mathrm D}/p_n^{\mathrm D}=\tan^2\theta$. The probabilities are therefore equal for all $n$ at $\theta=\pi/4$. Figure~\ref{fig:collective_battery}(a) shows $p_n^{\mathrm D}$ for $N=8$.

For comparison, we take $N$ independent charger--battery pairs $(C_j,B_j)$, each driven by the same $\Gamma_C(t)$ and $\Gamma_B(t)$ but coupled through its own single-pair decay channel $L_j=\sqrt{\Gamma_C}\,\sigma^-_{C_j}-\sqrt{\Gamma_B}\,\sigma^-_{B_j}$: different pairs do not share decay amplitudes, in contrast to the single collective channel $L_N$ of Eq.~\eqref{eq:2}. Along the corresponding dark-state family, the reduced battery state factorizes as
$
\rho_B^{\rm ind}
=
\bigotimes_{j=1}^{N}
\left(
\cos^2\theta\,|g_j\rangle\langle g_j|
+
\sin^2\theta\,|e_j\rangle\langle e_j|
\right),
$
where each cell is excited with probability $\sin^2\theta$; the excitation numbers are therefore binomially distributed,
$p_n^{\rm ind}=\binom{N}{n}\sin^{2n}\theta\cos^{2(N-n)}\theta$,
shown for $N=8$ in Fig.~\ref{fig:collective_battery}(b). At $\theta=\pi/4$, $p_n^{\mathrm D}=1/(N+1)$, while $p_n^{\rm ind}$ follows the binomial distribution. The $n$-excitation eigenspace of $H_B$ has dimension $\binom{N}{n}$. For independent charging, $p_n^{\rm ind}$ is the total probability of this eigenspace, shared equally among its $\binom{N}{n}$ product states. For collective charging, $p_n^{\mathrm D}$ is the probability of the single symmetric Dicke state $|n\rangle_B$ in the same eigenspace. Thus, for $0<\theta<\pi/2$, the independent battery state has rank $2^N$, whereas the collective battery state has rank $N+1$. These different spectra can give different passive energies and global ergotropies even at the same stored energy, as shown in Fig.~\ref{fig:collective_battery}(c).

From Eq.~\eqref{eq:reduced_battery}, the reduced state has at most $N+1$ nonzero eigenvalues $p_n^{\mathrm D}$. The battery Hamiltonian has a nondegenerate ground state and $N$ degenerate one-excitation states at energy $\omega_0$. The passive rearrangement places the largest eigenvalue $p_{\max}^{\mathrm D}=\max_n p_n^{\mathrm D}$ in the ground state and the remaining nonzero eigenvalues in the one-excitation manifold. The passive energy is therefore $E_{\rm pas}^{\mathrm D}=\omega_0(1-p_{\max}^{\mathrm D})$.

\begin{figure*}[!t]
\centering
\includegraphics[width=\textwidth]{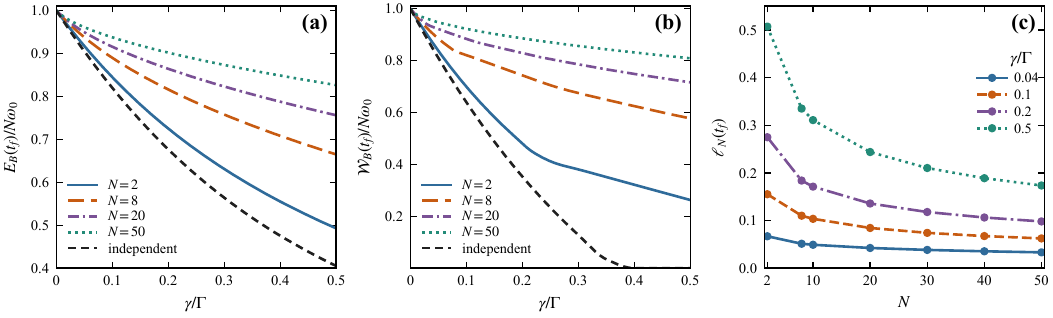}
\caption{Finite-rate collective charging and radiative loss. (a) Final normalized stored energy $E_B(t_f)/N\omega_0$ and (b) normalized global ergotropy $\mathcal W_B(t_f)/N\omega_0$ as functions of $\gamma/\Gamma$ for $N=2$, $8$, $20$, and $50$, calculated from Eq.~\eqref{eq:2} over $x\in[-7,7]$. The dashed curves show the corresponding independent-pair results. (c) Radiative loss per initial excitation $\ell_N(t_f)$ versus $N$ for the indicated sweep rates.} \label{fig:finite_rate}
\end{figure*}

Writing $r=\tan^2\theta$, Eq.~\eqref{eq:reduced_battery} gives $p_n^{\mathrm D}=r^n/\sum_{m=0}^{N}r^m$. The stored energy and global ergotropy are then
\begin{equation} \label{eq:generalN_battery}
\begin{aligned}
\frac{E_B^{\mathrm D}}{\omega_0}
&=
\sum_{n=0}^{N} n p_n^{\mathrm D}
=
\frac{
r\!\left[1-(N+1)r^N+Nr^{N+1}\right]
}{
(1-r)(1-r^{N+1})
},
\\
\frac{\mathcal{W}_B^{\mathrm D}}{\omega_0}
&=
\sum_{n=0}^{N} n p_n^{\mathrm D}
-1+p_{\max}^{\mathrm D}
\end{aligned}
\end{equation}
The $r\to1$ limit gives $E_B^{\mathrm D}=N\omega_0/2$, corresponding to half filling.

At $\theta=0$, the battery is in its ground state, with $E_B^{\mathrm D}=\mathcal W_B^{\mathrm D}=0$. At $\theta=\pi/2$, all battery qubits are excited and $E_B^{\mathrm D}=\mathcal W_B^{\mathrm D}=N\omega_0$. For $N>1$, the collective battery has nonzero global ergotropy even below half filling, where an independently charged battery with the same stored energy is passive. In this regime, $0<r<1$ and $p_{\max}^{\mathrm D}=p_0^{\mathrm D}$. Equation~\eqref{eq:generalN_battery} then gives
\begingroup
\begin{equation}
\frac{\mathcal W_B^{\mathrm D}}{\omega_0}
=
\sum_{n=2}^{N}(n-1)\,p_n^{\mathrm D}
>
0 .
\end{equation}
\endgroup
Thus, population inversion among the symmetric Dicke states is not required for nonzero global ergotropy.

For the collective dark states, $E_{\rm pas}^{\mathrm D}=\omega_0(1-p_{\max}^{\mathrm D})<\omega_0$. Since $p_{\max}^{\mathrm D}\geq1/(N+1)$, the normalized passive energy satisfies $E_{\rm pas}^{\mathrm D}/N\omega_0\leq1/(N+1)$. When the average stored energy per qubit remains fixed and nonzero, the passive energy becomes negligible compared with the stored energy, and the extractable fraction approaches 100\% as $N$ increases. This behavior is shown in Fig.~\ref{fig:collective_battery}(c).

For finite-rate charging, we choose $\theta(t)=\arctan e^x$, with $x=\gamma(t-\tau)$, where $\gamma$ is the sweep rate and $\tau$ sets the center of the sweep. The couplings are
\begingroup
\begin{equation} \label{eq:logistic}
\begin{aligned}
G_C&=G_0\sin\theta,
\qquad
G_B=G_0\cos\theta,
\\
\Gamma_C
&=
\frac{\Gamma}{2}
(1+\tanh x),
\qquad
\Gamma_B
=
\frac{\Gamma}{2}
(1-\tanh x),
\end{aligned}
\end{equation}
\endgroup
with $\Gamma=4G_0^2/\kappa$. For the logistic protocol, the slow-control requirement entering the cavity elimination reduces to $\gamma\ll\kappa$. The coefficients satisfy $\Gamma_C+\Gamma_B=\Gamma$.

For this protocol, $r=e^{2x}$. Substituting into Eq.~\eqref{eq:generalN_battery} gives the energy and ergotropy of the instantaneous dark states. Differentiating the energy with respect to time, using $dx/dt=\gamma$, we obtain
\begin{equation} \label{eq:logistic_branch}
\begin{aligned}
\frac{E_B^{\mathrm D}(x)}{\omega_0}
&=
\frac{N}{2}
+\frac{N+1}{2}\coth[(N+1)x]
-\frac{1}{2}\coth x,
\\
\frac{\mathcal W_B^{\mathrm D}(x)}{\omega_0}
&=
\frac{N}{2}
+\frac{N+1}{2}\coth[(N+1)x]
-\frac{1}{2}\coth x
-1
\\
&\quad
+
e^{N|x|}
\frac{\sinh |x|}{\sinh[(N+1)|x|]},
\\
\dot E_B^{\mathrm D}(x)
&=
\frac{\gamma\omega_0}{2}
\left[
\frac{1}{\sinh^2 x}
-\frac{(N+1)^2}{\sinh^2[(N+1)x]}
\right].
\end{aligned}
\end{equation}
At $x=0$, the rate of change of the dark-state energy is $\dot E_B^{\mathrm D}(0)=\gamma\omega_0N(N+2)/6$. Equation~\eqref{eq:logistic_branch} describes the instantaneous dark-state family. The actual finite-rate evolution generally deviates from this family and is determined by Eq.~\eqref{eq:2}.

The finite-rate evolution also remains within the symmetric subspace of each ensemble, since the initial state is symmetric and the collective jump operator preserves this symmetry. The jump operator lowers the total excitation number $Q=\hat n_C+\hat n_B$ by one, $[Q,L_N(t)]=-L_N(t)$. Starting from $Q=N$, the density matrix remains block diagonal in total excitation number. Tracing out the charger then gives $\rho_B(t)=\sum_{n=0}^{N}p_n(t)|n\rangle_B\langle n|$. Thus, the battery state has at most $N+1$ nonzero eigenvalues even at finite sweep rates, and its global ergotropy is $\mathcal W_B(t)=E_B(t)-\omega_0[1-p_{\max}(t)]$, where $p_{\max}(t)=\max_n p_n(t)$. The passive-energy bound therefore remains valid even when the system departs from the instantaneous dark state, whereas the stored energy depends on the actual finite-rate evolution.

We integrate Eq.~\eqref{eq:2} over $x\in[-7,7]$, i.e., from $t_i=\tau-7/\gamma$ to $t_f=\tau+7/\gamma$; the sweep center $\tau$ only shifts the time origin and does not affect the results for a fixed integration interval in $x$. The resulting normalized stored energy at the final time and normalized global ergotropy are shown in Figs.~\ref{fig:finite_rate}(a) and \ref{fig:finite_rate}(b). At fixed $\gamma/\Gamma$, both the normalized stored energy and global ergotropy increase with $N$ over the range studied. At $\gamma/\Gamma=0.5$, $E_B(t_f)/N\omega_0$ increases from 0.493 for $N=2$ to 0.826 for $N=50$, while $\mathcal W_B(t_f)/N\omega_0$ increases from 0.263 to 0.809. For $N=50$, these values correspond to a stored energy of approximately $41.3\omega_0$ and a passive energy of only $0.85\omega_0$. The independently charged battery is passive at this sweep rate.

To account for the reduction in stored energy, we define the fraction of initial excitations lost to the bath as $\ell_N(t_f)=1-\langle\hat n_C+\hat n_B\rangle_{t_f}/N$, and the fraction remaining in the charger as $R_N(t_f)=\langle\hat n_C\rangle_{t_f}/N$. Since $E_B(t_f)=\omega_0\langle\hat n_B\rangle_{t_f}$, these quantities satisfy
\begin{equation} \label{eq:balance}
1-\frac{E_B(t_f)}{N\omega_0}
=
\ell_N(t_f)+R_N(t_f).
\end{equation}
Figure~\ref{fig:finite_rate}(c) shows that radiative loss per initial excitation decreases with $N$ at the sweep rates considered. Equation~\eqref{eq:balance} separates the reduction in stored energy from full charge into radiative loss and excitations remaining in the charger. In the dilute limit, an additional bosonic counterdiabatic exchange enables exact following of the dark-state family at finite sweep rates.

\textit{Conclusion.---}
The comparison between collective and independent charging shows that equal stored energies need not yield equal global ergotropy. For $N>1$, the collectively charged battery has nonzero global ergotropy below half filling, whereas the independently charged reference battery with the same stored energy is passive. In our protocol, interference between collective decay amplitudes selects a tunable family of many-body dark states for transferring excitations from the charger to the battery. The collective dynamics preserves the symmetric Dicke subspace of the battery. Together with the degeneracy of the battery Hamiltonian, this keeps the passive energy of the entire battery below $\omega_0$, even when the stored energy grows linearly with $N$. When the stored energy per battery qubit remains fixed and nonzero, the ratio of global ergotropy to stored energy approaches unity as $N\to\infty$. The restriction to the symmetric subspace also holds at finite sweep rates. For the parameters studied, larger ensembles exhibit higher stored energy and global ergotropy per qubit, together with lower radiative loss per initial excitation.

\textit{Acknowledgments.---}
This work was supported by the National Natural Science Foundation of China (Grant No. 12275212) and the Natural Science Foundation of Shaanxi Province (Grant No. 2025JCYBQN055).

\setcounter{equation}{0}
\renewcommand{\theequation}{A\arabic{equation}}
\makeatletter
\def\widetext{%
  \par\ignorespaces
  \onecolumngrid
  \vskip8\p@
  \prep@math@patch
}%
\def\endwidetext{%
  \par
  \vskip6\p@
  \twocolumngrid\global\@ignoretrue
  \@endpetrue
}%
\makeatother
\begin{widetext}
\centering\textbf{End Matter}
\end{widetext}
\setlength{\abovedisplayskip}{6pt}
\setlength{\belowdisplayskip}{6pt}
\setlength{\abovedisplayshortskip}{6pt}
\setlength{\belowdisplayshortskip}{6pt}

\textit{Low-excitation limit.---}
The dark-state construction also applies to a fixed total excitation number $K\leq N$. Each ensemble contains $N$ qubits, and the dynamics is restricted to the fully symmetric Dicke subspace of each ensemble. The states with $K$ total excitations are spanned by $|K-n,n\rangle$, with $n=0,\ldots,K$, where $n$ excitations are in the battery and $K-n$ in the charger. A general state in this subspace is $|\psi_K\rangle=\sum_{n=0}^{K}d_n|K-n,n\rangle$. For the normalized symmetric Dicke states of an $N$-qubit ensemble, $S^-|m\rangle=\sqrt{m(N-m+1)}|m-1\rangle$. For $n=0,\ldots,K-1$, lowering the charger in $|K-n,n\rangle$ and lowering the battery in $|K-n-1,n+1\rangle$ lead to the same final state $|K-n-1,n\rangle$:
\begin{align}
&S_C^-|K-n,n\rangle
=
\sqrt{(K\!-\!n)(N\!-\!K\!+\!n\!+\!1)}
\,|K-n-1,n\rangle,
\\
&S_B^-|K-n-1,n+1\rangle
=
\sqrt{(n+1)(N-n)}
\,|K-n-1,n\rangle. \notag
\end{align}
Substituting these two contributions into $L_N|\psi_K\rangle$ gives
\begin{equation}
\label{eq:generalK_jump}
\begin{aligned}
L_N|\psi_K\rangle
&=
\sum_{n=0}^{K-1}\Bigl[
\sqrt{\Gamma_C}\,d_n
\sqrt{(K\!-\!n)(N\!-\!K\!+\!n\!+\!1)}
\\
&-\sqrt{\Gamma_B}\,d_{n+1}
\sqrt{(n+1)(N-n)}
\Bigr]
|K-n-1,n\rangle ,
\end{aligned}
\end{equation}
Since the final Dicke states $|K-n-1,n\rangle$ are orthogonal, the condition $L_N|\psi_K\rangle=0$ requires each corresponding coefficient in Eq.~\eqref{eq:generalK_jump} to vanish. With $\tan\theta=\sqrt{\Gamma_C/\Gamma_B}$, this gives
\begin{equation}
\frac{d_{n+1}}{d_n}
=
\tan\theta
\sqrt{
\frac{(K-n)(N-K+n+1)}
{(n+1)(N-n)}
} .
\label{eq:generalK_dark_recurrence}
\end{equation}
Equation~\eqref{eq:generalK_dark_recurrence} is exact within the symmetric subspace with $K$ total excitations at finite $N$. For $K=N$, it reduces to $d_{n+1}/d_n=\tan\theta$, recovering the constant adjacent-amplitude ratio of the dark state in the main text. In the limit $N\to\infty$ at fixed $K$, the factor $(N-K+n+1)/(N-n)$ approaches unity, and Eq.~\eqref{eq:generalK_dark_recurrence} reduces to $d_{n+1}/d_n=\tan\theta\sqrt{(K-n)/(n+1)}$. Iteration gives
\begin{equation}
\frac{d_n}{d_0}
=
\tan^{n}\theta
\sqrt{\binom Kn},
\end{equation}
Normalization gives $d_0=\cos^K\theta$, with a positive overall phase and $0\leq\theta<\pi/2$. The normalized dark state is therefore
\begin{equation}
|D_K^{\rm low}(\theta)\rangle
=
\sum_{n=0}^{K}
\sqrt{\binom Kn}\,
\cos^{K-n}\theta\,
\sin^n\theta\,
|K-n,n\rangle .
\label{eq:low_excitation_dark_state}
\end{equation}
As $\theta$ varies from $0$ to $\pi/2$, this state changes continuously from $|K,0\rangle$ to $|0,K\rangle$. Unlike the $K=N$ dark state, whose adjacent amplitudes have a constant ratio, the dilute-limit state contains the factor $\sqrt{\binom{K}{n}}$. The physical battery still consists of $N$ qubits, while the bosonic description applies to its collective excitations in the dilute limit.

In this limit, the collective spin excitations can be described using Holstein--Primakoff bosons $a_\alpha$, with $\alpha=C,B$ \cite{ref94}. For excitation numbers much smaller than $N$, the Dicke matrix elements give $S_\alpha^-\simeq\sqrt{N}\,a_\alpha$. The corresponding collective jump operator becomes $L_N\simeq L_{\rm low}=\sqrt{N}(\sqrt{\Gamma_C}\,a_C-\sqrt{\Gamma_B}\,a_B)$. By the binomial theorem, Eq.~\eqref{eq:low_excitation_dark_state} can equivalently be written as $|D_K^{\rm low}(\theta)\rangle=[d^\dagger(\theta)]^K|0,0\rangle/\sqrt{K!}$, with $d^\dagger(\theta)=\cos\theta\,a_C^\dagger+\sin\theta\,a_B^\dagger$; this is the $K$-excitation Fock state of the collective dark mode. Changing $\theta$ rotates the two collective bosonic modes. The corresponding finite-rate generator is \cite{ref95,ref96}
\begin{equation}
H_{\rm cd}^{\rm low}(t)
=
i\dot\theta(t)
\left(
a_B^\dagger a_C-a_C^\dagger a_B
\right).
\label{eq:low_excitation_cd}
\end{equation}
For $G=a_B^\dagger a_C-a_C^\dagger a_B$, $[G,d^\dagger(\theta)]=\partial_\theta d^\dagger(\theta)$ and $G|0,0\rangle=0$. Equation~\eqref{eq:low_excitation_cd} therefore gives $H_{\rm cd}^{\rm low}|D_K^{\rm low}\rangle=i\partial_t|D_K^{\rm low}\rangle$. At the same time, $L_{\rm low}|D_K^{\rm low}\rangle=0$ because $\sqrt{\Gamma_C}\cos\theta=\sqrt{\Gamma_B}\sin\theta$. Thus $\rho_D(t)=|D_K^{\rm low}(t)\rangle\langle D_K^{\rm low}(t)|$ exactly satisfies the bosonic low-excitation master equation
\begin{equation}
\dot\rho
=
-i[H_{\rm cd}^{\rm low},\rho]
+
\mathcal D[L_{\rm low}]\rho .
\label{eq:low_excitation_master}
\end{equation}
\begin{figure}[t]
\centering
\includegraphics[width=0.88\columnwidth]{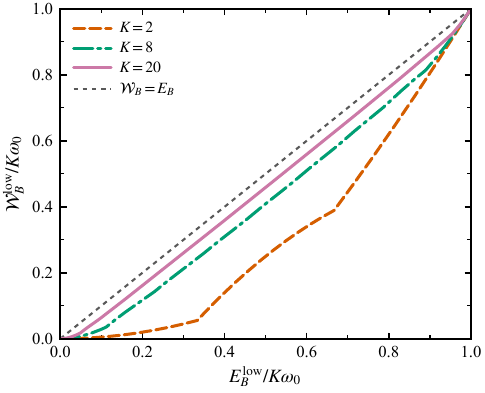}
\caption{Extractable work in the low-excitation limit. Global ergotropy $\mathcal W_B^{\rm low}/K\omega_0$ versus stored energy $E_B^{\rm low}/K\omega_0$ for $K=2$, $8$, and $20$ in the dilute limit $N\rightarrow\infty$ at fixed $K$. The dotted diagonal denotes $\mathcal W_B^{\mathrm{low}}=E_B^{\mathrm{low}}$.} \label{fig:low_excitation_ergotropy}
\end{figure}
Equation~\eqref{eq:low_excitation_cd} describes a phase-controlled beam-splitter interaction between the two collective bosonic modes, with an exchange phase of $\pi/2$. Such exchange interactions are available in circuit QED~\cite{ref39,ref35,ref37}. For the finite-spin dark states with $K=N$, however, a uniform collective-spin exchange does not generally generate $\partial_\theta|D_N\rangle$ because of the different Dicke matrix elements.

The charger states associated with different $n$ are orthogonal, ${}_C\langle K-m|K-n\rangle_C=\delta_{mn}$, so tracing out the charger removes all $m\neq n$ cross terms, with $q=\sin^2\theta$:
\begin{equation}
\rho_B^{\rm low}
=
\sum_{n=0}^{K}
p_n^{\rm low}|n\rangle_B\langle n|,
\quad
p_n^{\rm low}
=
\binom Kn q^n(1-q)^{K-n} .
\label{eq:low_excitation_rhoB}
\end{equation}
Here $p_n^{\rm low}$ is the probability that the battery ensemble contains $n$ total excitations, while $|n\rangle_B$ is the symmetric $n$-excitation Dicke state distributed over the $N$ physical battery qubits. The reduced battery spectrum is binomial. The spectral structure differs from the independent-pair case in Fig.~\ref{fig:collective_battery}(b): each excitation sector contributes only one symmetric Dicke eigenvector, so $\rho_B^{\rm low}$ has at most $K+1$ nonzero eigenvalues. Using $n\binom Kn=K\binom{K-1}{n-1}$, the binomial mean is $\langle\hat n_B\rangle=Kq$.
Work extraction is evaluated with the physical $N$-qubit Hamiltonian $H_B=\omega_0\hat n_B$, whose lowest levels consist of one ground state and an $N$-fold degenerate one-excitation manifold. Since $\rho_B^{\rm low}$ has at most $K+1\leq N+1$ nonzero eigenvalues, passive rearrangement places $p_{\max}^{\rm low}=\max_{0\le n\le K}p_n^{\rm low}$ on the ground state and the remaining positive eigenvalues in the one-excitation manifold. These remaining eigenvalues sum to $1-p_{\max}^{\rm low}$, giving $E_{\rm pas}^{\rm low}=\omega_0(1-p_{\max}^{\rm low})$. With $\mathcal W_B=E_B-E_{\rm pas}$, the stored energy and global ergotropy are
\begin{equation}
E_B^{\rm low}=K\omega_0q,
\qquad
\mathcal W_B^{\rm low}
=
K\omega_0q
-
\omega_0(1-p_{\max}^{\rm low}).
\label{eq:low_excitation_energy_ergotropy}
\end{equation}
Permutation symmetry gives $\langle\hat n_B\rangle=Np_e^{(1)}$, where $p_e^{(1)}$ is the excited-state population of any single battery qubit. Hence $p_e^{(1)}=Kq/N$, and the single-cell state is
$
\rho_j
=
(1-p_e^{(1)})|g\rangle\langle g|
+
p_e^{(1)}|e\rangle\langle e| .
$
At fixed $K$, $p_e^{(1)}\to0$ as $N\to\infty$, so the single-cell state is passive. At the collective level, $E_B^{\rm low}-\mathcal W_B^{\rm low}=\omega_0(1-p_{\max}^{\rm low})<\omega_0$. Relative to the full battery capacity, $E_B^{\rm low}/N\omega_0=(K/N)q$ and $\mathcal W_B^{\rm low}/N\omega_0\leq K/N$. Figure~\ref{fig:low_excitation_ergotropy} therefore uses $K\omega_0$, the energy available in the fixed-$K$ sector, as the normalization. Because the exchange amplitude in Eq.~\eqref{eq:low_excitation_cd} equals $\dot\theta(t)$, the same construction applies to any smooth trajectory $\theta(t)$. For the logistic protocol, $q=[1+\tanh x]/2$ and $\dot\theta=(\gamma/2)\operatorname{sech}x$. Increasing $\gamma$ shortens the traversal time and increases the required exchange amplitude; the $\mathcal W_B(E_B)$ trajectory is unchanged as long as the bad-cavity condition $\gamma\ll\kappa$ remains satisfied.

\end{document}